\documentclass[proceedings]{JHEP3}

\PrHEP{PrHEP hep2001}                   
\conference{International Europhysics Conference on HEP}                

\usepackage{color}      
\usepackage{graphicx}   
\DeclareGraphicsExtensions{.eps,.ps}
\usepackage{layout}
\usepackage{epsfig}

\newcommand\sele{\mbox{$\widetilde{e}$}}
\newcommand\squa{\mbox{$\widetilde{q}$}}
\newcommand\supq{\mbox{$\widetilde{u}$}}
\newcommand\scha{\mbox{$\widetilde{c}$}}
\newcommand\sbot{\mbox{$\widetilde{b}$}}
\newcommand\stopq{\mbox{$\widetilde{t}$}}

\title{Search for \boldmath{$R_{p}$}-violating SUSY and excited
fermions at HERA}

\author{\speaker{Christian Schwanenberger}                 
        \thanks{Talk given at the International Europhysics Conference on High
                           Energy Physics, July 12-18, 2001, Budapest,
        Hungary.}\\                                         
        Deutsches Elektronen-Synchrotron DESY, Notkestra{\ss}e 85, D-22607
        Hamburg, Germany\\                                 
        E-mail: \email{schwanen@mail.desy.de}}             

\author{on behalf of the H1 and ZEUS collaborations}

\abstract{Recent results from searches for physics beyond the
Standard Model (SM)  
in $e^\pm$-proton collisions at a center of mass energy of 300 --
318 GeV at HERA are presented. Searches for excitations of fermions and for the
production of squarks in R-parity-violating Supersymmetry (SUSY) are
reviewed.}

\begin{document}


\section{Introduction}

At the HERA collider, electrons (positrons) and protons collide at a
center of mass energy of about $\sqrt{s}= 318$ GeV (300 GeV before
1998). Since 1994, integrated luminosities of about ${\cal L} = 110\ {\rm
pb}^{-1}$ in $e^+ p$ and ${\cal L} = 15\ {\rm pb}^{-1}$ in $e^- p$
scattering have been collected by each of the two experiments H1 and
ZEUS. These 
data are used to search for new physics beyond the SM. Recent results on
searches for excited 
fermions and for R-parity-violating SUSY are presented in this
contribution.


\section{Excited fermions}

Fermionic substructure as predicted in compositeness models naturally leads
to excitations of fermions. 
At HERA, such excited fermions could be singly produced via the t-channel
exchange of a gauge boson, and would subsequently decay into a SM fermion
and a boson.
In the 1994--97/1999--2000 $e^+ p$ and the
1998--99 $e^- p$ data, both collaborations have found no evidence for
a deviation from the SM due to the production of an excited
electron\footnote{If not
particularly emphasized, {\it electron} can mean either an
electron or a positron.}, neutrino or quark, which decay into
$\gamma$, $Z$ or $W$ and a SM fermion, followed by the subsequent decay of
the boson into 
$e$, $\mu$, $\nu$ or hadrons. 

\FIGURE{
   \mbox{\epsfig{file=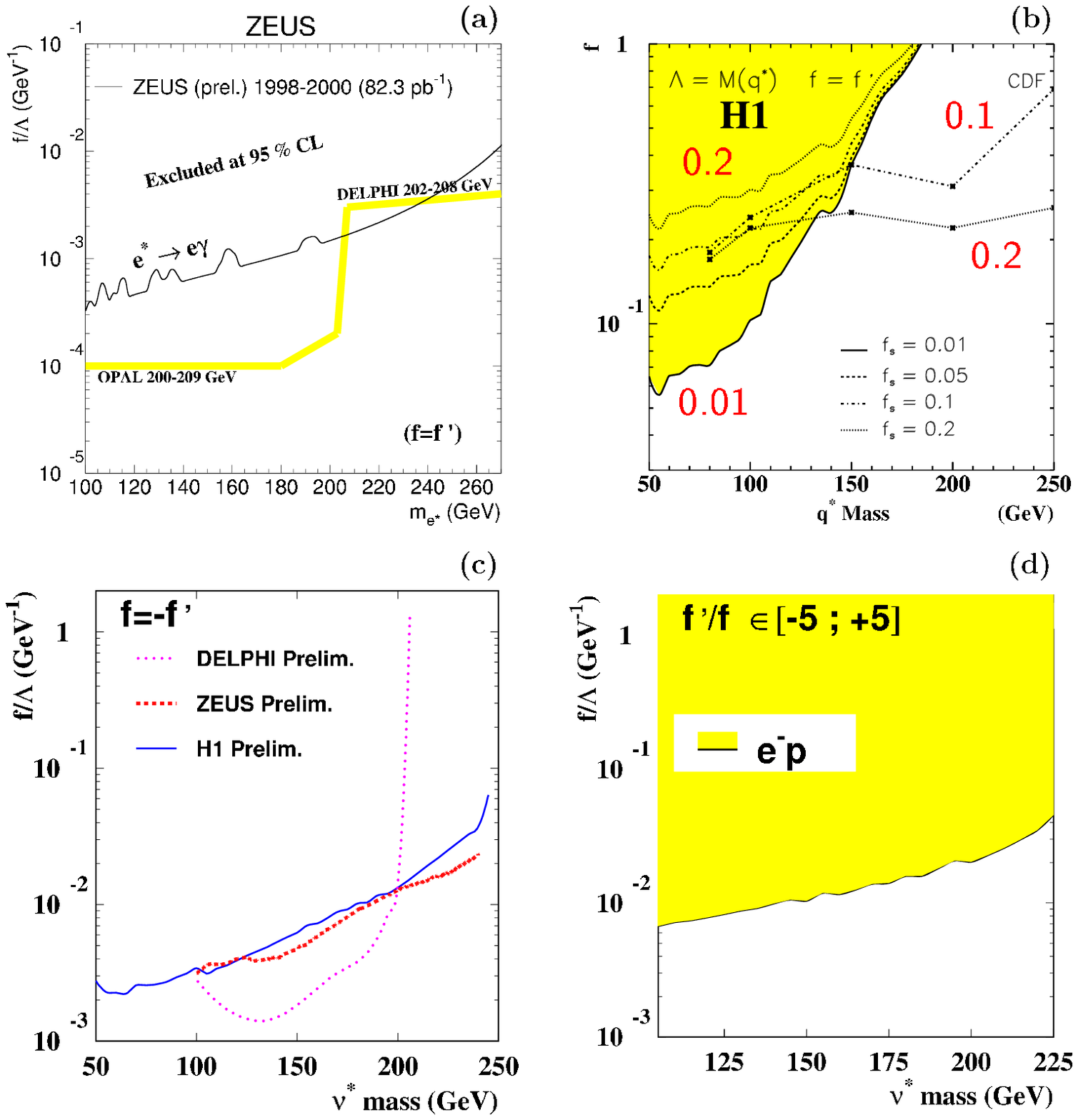,width=13.cm,bbllx=65,bblly=570,bburx=535,bbury=90,clip=}}
\vspace{12.85cm}
\caption{{(a) ZEUS upper limits on $f/\Lambda$, under the
  assumption $f=f'$, as a function of the excited electron mass are
  compared to those from OPAL and DELPHI. (b)
  H1 upper limits on $f$,
  under the assumptions $f=f'$ and $\Lambda=M(q^*)$ for different values of
  $f_s$, as a function of the excited quark mass are compared to CDF
  results. 
  (c) ZEUS and H1 upper limits on $f/\Lambda$, under the
  assumption $f=-f'$ as a function of the excited neutrino mass, are
  compared to those of DELPHI. (d) H1 upper limits on $f/\Lambda$,
  as a function of the excited neutrino mass, are
  presented for arbitrary ratios $f'/f$.}}\label{fig:excfer}
}

To determine the experimental sensitivity, the data have been interpreted in
the framework of a compositeness 
model where the excited fermion $f^*$ is a composite
of a scalar and a spin
$\frac{1}{2}$ constituent~\cite{compmod}. If $f^*$ carries spin
and isospin $\frac{1}{2}$, the effective Lagrangian
reads \cite{exclag}
\begin{eqnarray}
  {\cal L}_{eff} = \frac{1}{2{\Lambda}}\  { {\bar{F}_R}^*}\ 
  \sigma^{\mu\nu} \,(\, g f\, \frac{\tau^{a}}{2}\, 
   {W^a_{\mu\nu}}
  +  g' f' \,\frac{Y}{2}\,  B_{\mu\nu}
  +  g_s  f_s \,\frac{\lambda_a}{2}\,
   G^a_{\mu\nu}\,)\,  F_L  \ \ + \ \ h.c. \,\, ,
\end{eqnarray}
with the standard electroweak and strong couplings denoted by $g$, $g'$ and $g_s$ and the SM
gauge field strength tensors ${W^a_{\mu\nu}}$ (SU(2)), $B_{\mu\nu}$ (U(1)) and
$G^a_{\mu\nu}$ ({$\rm SU(3)_C$}). ${\tau^{a}}$, $Y$ and ${\lambda_a}$ are
the Pauli matrices, the weak hypercharge operator and the Gell-Mann
matrices, respectively. The new couplings between the weak
isodoublets $F^*_R$, $F^*_L$ of the excited fermion fields and the SM
fermion fields $F_R$, $F_L$ are modified by $f$, $f'$ and
$f_s$. The compositeness scale is $\Lambda$. The coupling between e.g. 
a photon,
an excited fermion and a SM fermion is given by 
   $c_{\gamma f^* f} = \frac{1}{2}\ ( {f}\ I_3 + {f'}\ 
   \frac{Y}{2} )$,
where $I_3$ is the third component of the isospin of the fermion.
Assuming relations between the new couplings, upper limits on $f$ (or
$f/\Lambda$) at the 95\%\
confidence level (CL) have been obtained. 

For excited electrons, the ZEUS upper limits have been improved recently by a
factor of 2 -- 3 and are competitive with those from LEP for larger masses
$m_{e^*}$~\cite{zeusexcfer} as can be inferred from Figure~\ref{fig:excfer}a. 
Figure~\ref{fig:excfer}b shows
that, for small values of $f_s$, the H1 upper limits on $f$ ($f=f'$;
$\Lambda=M(q^*)$) for
excited quarks are more restrictive than those of the TeVatron for not too
large masses $M(q^*)$. The HERA results obtained on $q^*$ production via the
electroweak couplings $f$ and $f'$ are complementary to the results
obtained at the 
TeVatron, where $q^*$ production
occurs via the strong coupling $f_s$.

From Figure~\ref{fig:excfer}c it can be seen that the
H1~\cite{h1excfer} and 
ZEUS~\cite{zeusexcfer2} upper limits on $f/\Lambda$ 
($f=-f'$) for excited neutrinos deliver a unique sensitivity beyond the LEP
center of mass energy.\footnote{A similar picture can be found for the upper limits
for excited quarks.} 
In addition,
H1 has obtained less model-dependent upper limits on $f/\Lambda$ for
arbitrary ratios $f'/f$~\cite{h1excneu}. This is presented in
Figure~\ref{fig:excfer}d. Since $\nu^*$ results were obtained from
a relatively small amount of $e^- p$ \footnote{In this case, $e^- p$
scattering is superior to
$e^+ p$ because the dominant valence quark interaction is helicity
suppressed and has smaller parton densities in $e^+ p$ ($d$) compared to
$e^- p$ ($u$).} data, new $e^- p$
luminosity from 2001 (see Section 4) will give substantially higher
sensitivity to excited neutrinos and are
expected to improve on the LEP limits, even for masses below
200 GeV.


\section{Squarks in R-Parity-violating SUSY}

In the Minimal Supersymmetric Standard Model (MSSM), there is a
multiplicative quantum number 
called R-Parity which is defined by $R_p \equiv (-1)^{3B+L+2S}$, 
with the baryon number is denoted $B$, the lepton number $L$ and the spin quantum
number $S$. For SM particles (sparticles, the supersymmetric partners of the
particles), this quantum number acquires the value $R_p=1$ ($R_p=-1$). The
$R_p$-conserving MSSM has been 
investigated at HERA in the associated \sele\,\squa\ (selectron-squark)
production \cite{rpc}.

\FIGURE{
   \mbox{\epsfig{file=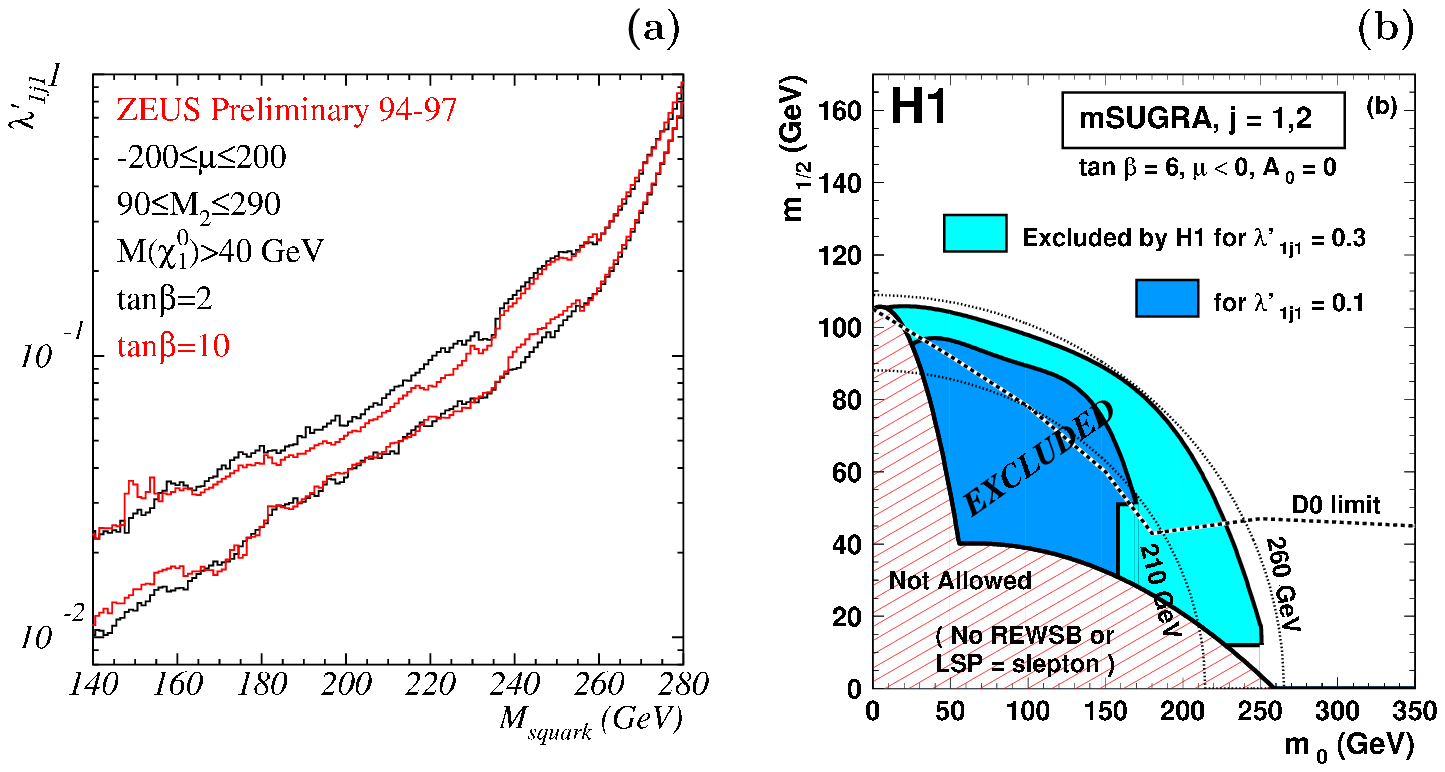,width=12.7cm,bbllx=70,bblly=775,bburx=500,bbury=550,clip=}}
   \vspace{6.4cm}
      {\caption
 {(a) as a function of the squark mass, for two values of
 $\tan\beta$, the upper and the lower curves give the maximal and minimal
 values for the upper limits on the R-Parity-violating SUSY coupling
 $\lambda'$ (unconstrained MSSM) when varying $\mu$ and $M_2$. (b)
 for $\tan\beta=6$ and two values of $\lambda'$ (mSUGRA scheme), the
 excluded region in the ($m_0$,$m_{1/2}$) plane is depicted. The region
  below the 
 dashed curve is ruled out by the D0 experiment. The contours for equal
 squark masses are also shown.}\label{fig:rpv}
}}  

However, the most general supersymmetric and gauge-invariant superpotential
contains three 
R-Parity-violating Yukawa coupling terms, two operators mediating $L$
violation and one $B$ violation~\cite{barger}. Of particular interest for
HERA are the terms
$\lambda'_{ijk} L_i Q_j \bar{D}_k$ with $L_i$ ($Q_j$) the
lepton (quark) $SU(2)_L$ doublet superfields and $\bar{D}_k$ the down-quark
$SU(2)_L$ singlet superfields~\cite{rpv}. The R-Parity-violating dimensionless 
coupling constants are called ${\lambda'_{ijk}}$~, 
where $i$, $j$, $k$ denote the generation.
For non-vanishing ${\lambda'_{1jk}}$~, the resonant production of single
squarks is possible in $ep$ scattering. At HERA with an $e^+$ beam,
predominantly the ${\lambda'_{1j1}}$ couplings are probed.

Both H1 and ZEUS have searched for squark production 
using $e^+ p$ data from 1994--97 at a center-of-mass energy
$\sqrt{s}=300$ GeV 
distinguishing between R-Parity-violating $\ell q$ ($\ell = e, \nu$) decays
leading to 
lepton + jet
topologies 
and decays into the superpartner of a gauge boson (gaugino) leading to lepton(s) + multi-jets topologies in the final
state. 
The gaugino can decay to a lepton
($e$ or $\nu_e$) and two quarks with the same $\lambda'$ coupling. It also
can decay to a
lighter gaugino and two fermions.
H1 and ZEUS found 
no evidence for squark production. The
results translate to constraints in SUSY parameter space in
various scenarios. 

Limits on R-Parity-violating couplings have been derived in the
unconstrained MSSM where the sfermion and gaugino sector
are assumed to be unrelated. The sfermion masses are free parameters. A scan in
the SUSY parameter space ($M_2$,$\mu$), $M_2$ being the soft
SUSY breaking mass term and $\mu$ the mixing mass term for 2 Higgs doublets,
has been performed 
for two different values of $\tan\beta$, the ratio of the vacuum
expectation values of the two neutral 
scalar Higgs fields. As a result, Figure~\ref{fig:rpv}a presents
the ZEUS upper limits on $\lambda'_{1j1}$ as a function of the squark
mass~\cite{zeusrpv}. They are widely SUSY parameter
independent. For \scha\ and \stopq\ production the constraints on
$\lambda'_{1j1}$ are more stringent than the indirect bounds from atomic parity
violation measurements~\cite{apv,h1rpv}.

H1 also sets limits~\cite{h1rpv} in constrained models where the sfermion
and gaugino
sector are related at the GUT scale and evolved down to the electroweak
scale by the Renormalization Group Equations. Minimal
Supergravity (mSUGRA) models assume, in addition, that the electroweak symmetry
breaking is driven by radiative corrections. This leads to 5 independent
parameters: $\tan\beta$, the sign of $\mu$, the common scalar mass $m_0$,
the common gaugino mass $m_{1/2}$ and the common trilinear coupling at the
GUT scale $A_0$. Figure~\ref{fig:rpv}b shows the H1 limits for fixed
values
of $\lambda'$ and $\tan\beta$ in the ($m_0$,$m_{1/2}$)
plane for \supq, \scha\ 
production. Here, for reasonably large values of the Yukawa coupling (even
as small as 0.1), H1 probes a domain which is not ruled out by the
($\lambda'$ independent) TeVatron
limits. For $\lambda'_{1j1}$ values
of the electromagnetic strength ($\approx 0.3$), squark masses below 260 GeV
are ruled out. However, it should be noted that 
LEP has
the highest sensitivity for \supq, \scha\ production, while for \stopq\ production for intermediate values of $m_0$ the H1 sensitivity is comparable with
LEP. Note that the results are from 1994--97
data, and there are further data taken since 1998 at a higher
center-of-mass energy, 318 GeV. There are also decays not yet 
investigated, for example $\stopq \rightarrow \sbot + W^+$, which can
produce an event 
topology with an isolated lepton and missing transverse momentum,
accompanied by a hadronic system with large transverse momentum. It is
noteworthy that H1 observes an excess above the SM prediction in this
topology~\cite{isolept}.


\section{Summary and prospects}

In the HERA searches for R-Parity-violating SUSY and for excited fermions,
no evidence for physics beyond the SM has been found. New constraints on
squark production in R-Parity-violating SUSY and for excitations of
electrons, neutrinos and quarks have been obtained. The HERA limits
complement and are
competitive with those of the LEP and TeVatron
searches. After the HERA detector and luminosity upgrade in autumn 2001,
HERA2 will accumulate approximately 1 $\rm fb^{-1}$ of luminosity. Thus,
a remarkably enhanced
discovery potential will be available for the analyses presented here. In
particular, the
polarized $e^\pm$ beams will lead to an increased squark production cross
section and sensitivity.


\section{Acknowledgements}

I wish to thank my colleagues from H1 and ZEUS collaborations for their
help in preparing this talk.

%


\end{document}